\documentclass[pra,twocolumn,preprintnumbers,amsmath,amssymb,superscriptaddress,showpacs,longbibliography]{revtex4-2}
\renewcommand{\thispagestyle}[1]{}
\usepackage[]{newtxtext} 
\usepackage[subscriptcorrection,nosymbolsc,smallerops,bigdelims]{newtxmath} 
\DeclareMathAlphabet{\mathcal}{OMS}{cmsy}{m}{n} 
\DeclareMathAlphabet{\mathbcal}{OMS}{cmsy}{b}{n} 
\usepackage{bm}

\usepackage{diagbox}

\usepackage[utf8]{inputenc}
\usepackage[T1]{fontenc}

\usepackage[]{graphicx}
\usepackage{latexsym}
\usepackage{color}
\usepackage{mathtools}
\usepackage[section]{placeins}
\usepackage[]{xcolor}
\usepackage{bm}
\usepackage{bbold}
\usepackage{multirow}
\usepackage{siunitx}
\usepackage{hyperref}
\hypersetup{
	colorlinks,
	linkcolor={blue!90!black},
	citecolor={blue!90!black},
	urlcolor=	{blue!90!black}
}
\let\oldeqref\eqref
\renewcommand*{\eqref}[1]{%
	\hyperref[#1]{\oldeqref{#1}}%
}

\usepackage{floatrow}
\usepackage{placeins}

\newcommand{\Tr}{\mathrm{Tr}}

\mathchardef\mhyphen="2D

\DeclarePairedDelimiterX{\comm}[2]{\lbrack}{\rbrack}{#1, #2}
\DeclarePairedDelimiter\ket{\lvert}{\rangle}
\DeclarePairedDelimiter\bra{\langle}{\rvert}
\DeclarePairedDelimiterX{\braket}[2]{\langle}{\rangle}{#1\delimsize\vert #2}
\DeclarePairedDelimiterX{\ketbra}[2]{\rvert}{\lvert}{#1 \delimsize\rangle\!\delimsize\langle #2}
\DeclarePairedDelimiterX{\matrixel}[3]{\langle}{\rangle}{#1 \delimsize\vert #2 \delimsize\vert #3}

\newcommand{\eqnref}[1]{Eq.~\eqref{#1}}

\newcommand{\figref}[1]{Fig.~\ref{#1}}
\newcommand{\ifigref}[1]{Figure~\ref{#1}}
\newcommand{\subfigref}[2]{Fig.~\hyperref[#1]{\ref*{#1}(#2)}}
\newcommand{\subfigsref}[3]{Figs.~\hyperref[#1]{\ref*{#1}(#2)}-\hyperref[#1]{\ref*{#1}(#3)}}

\usepackage[low-sup]{subdepth}
\usepackage{ragged2e}

\definecolor{cbred}{HTML}{e31a1c}
\definecolor{cbgreen}{HTML}{33a02c}
\definecolor{cbblue}{HTML}{176aa7}
\definecolor{cborange}{HTML}{ff7f00}
\definecolor{cbviolet}{HTML}{6a3d9a}
\definecolor{cbbrown}{HTML}{b15928}

\definecolor{cblred}{HTML}{fb9a99}
\definecolor{cblgreen}{HTML}{b2df8a}
\definecolor{cblblue}{HTML}{a6cee3}
\definecolor{cblorange}{HTML}{fdbf6f}
\definecolor{cblviolet}{HTML}{cab2d6}
\definecolor{cblbrown}{HTML}{ffff99}

\begin{document}

\title{Direct access to the initial polarization of ${}^{13}C$ nuclei by measuring coherence evolution of an nitrogen-vacancy center spin qubit}

\author{Mateusz Kuniej}
\affiliation{Institute of Theoretical Physics, Faculty of Fundamental Problems of 
	Technology, Wroc{\l}aw University of Science and Technology,
	50-370 Wroc{\l}aw, Poland}
\author{Katarzyna Roszak}
\affiliation{FZU - Institute of Physics of the Czech Academy of Sciences, 182 00 Prague, Czech Republic}

\date{\today}

\begin{abstract}
We introduce a method for the measurement of the lower bound on the initial polarization of spinful nuclei in a diamond by following the coherence evolution of an NV center spin qubit after a simple scheme is operated on the qubit to facilitate the transfer of information from the environment into the qubit state. Current polarization measurement techniques are challenging to implement due to the need for direct access to the environment. In our method, information is obtained by measuring the difference of the evolution of the qubit coherence resulting from preparation phase when the environment evolution is conditional on the qubit pointer state. We find that the method does not depend strongly on the applied magnetic field, but rather on the number of spinfull nuclei that lead to decoherence, and gives a reasonable estimate if the environment is polarized. The key advantage of this approach is its simplicity and minimal experimental requirements, allowing the inference of initial nuclear polarizations without direct access to the environment. We demonstrate the efficacy of this method using a simulated environment of up to fifteen randomly placed nuclear spins.
\end{abstract}

\maketitle
	
	
\section{Introduction \label{sec1}}
	
Nitrogen-vacancy (NV) centers in diamond \cite{Doherty2013,Wood2018,Awschalom2018,Tchebotareva2019} have recently attracted much attention due to their wide application potential in highly sensitive magnetometry \cite{Maze2008, Taylor2008, Hall2009, deLange2011}. For efficient operation, they require minimization of the decoherence resulting from interacting with imperfections within the diamond lattice. There are two dominating decoherence mechanisms for an NV center spin. Both are facilitated by the hyperfine interaction, either with spinful ${}^{13}C$ carbon defects in the lattice \cite{Zhao2012,Kwiatkowski2018}, or via a coupling to the P1 centers \cite{Bussandri2024, Park2022}. The concentration of the P1 centers can be significantly reduced for artificial diamonds \cite{Chen2023, Bauch2018}, thus we concentrate on the ${}^{13}C$ carbon isotopes as the source of decoherence.
 
The ${}^{13}C$ isotopes are rare in diamond and thus the spin qubit interacts with a relatively small number of environmental spins \cite{Maze2008b,Zhao2011}
which are randomly distributed within the crystal lattice. This leads to extremely long decoherence times \cite{Kennedy2003,Gaebel2006,deLange2011}, but even longer coherence times are necessary for effective use in ultrasensitive magnetometry. Consequently, several techniques such as dynamical decoupling (DD) \cite{Viola1999,deLange2010,Ryan2010,Naydenov2011} and dynamical nuclear polarization (DNP) \cite{London2013,Fischer2013,Pagliero2018,Wunderlich2017, Scheuer2017,Poggiali2017,Hovav2018} have to be employed. 

In recent years, different techniques have been used to polarize the spins of carbon isotopes, which leads to the reduction of qubit decoherence. The most promising technique uses an optically pumped NV center \cite{London2013, Fischer2013, Jacques2009, King2015, Alvarez2015}. However, determining the polarization of the nuclear spins after optical pumping is difficult. The most accurate methods require direct access to the environment, which can be experimentally challenging. However, such methods as polarization readout by polarization inversion (PROPI), optically detected magnetic resonance (ODMR), or exploiting the Fourier-transformed nuclear magnetic resonance spectrum allow finding the information about the state the environment \cite{Scheuer2017, Jacques2009, King2015}. In addition to that, it is possible to find the joint effect that spins have on the NV ensemble by summing the individual spin polarizations using an echo-based technique \cite{London2015}. Recently, a method has been developed that employs a simple measurement of NV coherence without the echo, allowing the estimation of the upper bound of the mean polarization of the nuclear spins of ${}^{13}C$ atoms \cite{Kuniej2025}. Since the evolution of the NV center spin qubit depends on each of the initial polarizations of environmental spins, indirect access to the nuclear polarization via qubit coherence is possible. 

Our approach is meant to facilitate the extraction of information about the initial polarization of nuclear ${}^{13}C$ spins by implementing a series of straightforward operations and measurements on a single NV center spin qubit that expedite the transfer of information about the environment into 
the qubit state. It serves as a simple way to estimate the level of success achieved in polarizing the environment, as it provides the lower bound on the average of the initial polarizations of the ${}^{13}C$ nuclear spins.
As in the method of Ref.~\cite{Kuniej2025}, it does not require the environment to be directly accessed. Instead, a protocol is performed solely on the NV-center spin qubit to transfer the information about the state of the environment into the qubit subspace and allow for readout of that information by measurement performed on the qubit.
    
We address the question of how to estimate the initial polarization of the nuclear spin environment using a scheme designed to detect qubit-environment entanglement (QEE) in pure dephasing evolutions \cite{Roszak2019}. Due to the large value of the ratio of electronic and nuclear gyromagnetic factors for the NV-center and ${}^{13}C$ nuclei, environment-induced transitions between different qubit states tend to be suppressed \cite{Zhao2012} and the NV-center decoherence is limited to pure dephasing. Thus, the possibility of QEE generation during joint qubit-environment evolution for this system can be qualified on the basis of the initial state of the environment \cite{Roszak2020}. A fully mixed state of the environment will never entangle with the qubit, but all at least partially polarized states lead to the formation of QEE \cite{Strzalka2020}. Moreover, the maximum amount of entanglement that can be obtained is proportional to the initial purity of the environmental state and thus proportional to the initial polarization of the state \cite{Strzalka2020}. 

Although such entanglement cannot be measured directly, there exist schemes for the detection of QEE in pure dephasing evolutions that involve only operations and measurements on the qubit \cite{Roszak2019,Rzepkowski2020,Strzalka2021,Strzalka2024}. They rely on the fact that the presence of QEE in such evolutions manifests itself in the state of the environment and can therefore have an effect on the qubit evolution through backaction. For pure decoherence in general, the schemes provide measurable entanglement witnesses \cite{Terhal2000,Lewenstein2000,Guhne2002,Barbieri2003}, but in specific situations, such as an NV center spin qubit interacting with a given environment (a diamond lattice where the locations of the ${}^{13}C$ nuclear spins are set), the measured signal can be proportional to the amount of QEE. 

We demonstrate how to extract information about the lower bound on the average nuclear polarization of the environment. To this end we employ a scheme, where firstly information about the initial state of the qubit is transferred into the state of the environment during a preparation phase, when the qubit is initialized in one of the spin eigenstates. The state of the nuclear environment after this phase is dependent on the concrete eigenstate, and when a superposition is excited on the qubit directly afterward, the decoherence is different for different qubit states in the preparation phase. Thus, information about the reaction of the environment to the qubit eigenstate can be read out from the qubit evolution, and since this depends on the actual polarization of the environment, a comparison between the resulting curves of qubit decoherence allows for the extraction of information about polarization. 

The correspondence between the decoherence and initial polarization depends on many fundamentally unknown variables, most notably on the coupling strengths between each randomly located spinfull carbon isotope
environmental spin and the qubit. Nevertheless, it is possible to find a lower bound on the initial polarization using two different approaches in the analysis of the decoherence curves. The simplest case requires only the measurement of the maximum difference between the differently prepared decoherence curves, while the second also uses the time at which the maximum is measured and the value of the applied magnetic field. The bound obtained via the time-dependent method is significantly more precise, especially considering that the additional requirements are not particularly stringent. 
We supplement our findings with simulations involving up to $15$ environmental nuclei and show the reliability of the estimates as it depends on the number of nuclei, magnetic field, and initial environmental polarization.

The paper is organized as follows. In Sec.~\ref{sec:sys&H} we discuss the evolution of the spin qubit in the presence of a nuclear environment. In Sec.~\ref{sec:ent} we introduce the scheme which is used to transfer information about the initial state of the environment into the qubit state and then read it out. In Sec.~\ref{sec:results} we present the methods for the estimation of the initial polarization of the environment by an analysis of the qubit decoherence evolution and show results for a specific realization of  a single nuclear environment. Sec.~\ref{sec:conclusion} concludes this paper.
	

\begin{table}[]
    \centering
    \begin{ruledtabular}
        \begin{tabular}{c c c c c}
            $k$&$r_k$~(nm)&$\mathbb{A}^{\mathrm{z},\mathrm{x}}_k$~(1/$\mu$s)&$\mathbb{A}^{\mathrm{z},\mathrm{y}}_k$~(1/$\mu$s)&$\mathbb{A}^{\mathrm{z},\mathrm{z}}_k$~(1/$\mu$s)  \\\hline
            $1$&0.50442 &1.37617   & 0        &0.973096\\
            $2$&0.56396 & 0.492352 &-0.511667 &-0.417774\\
            $3$&0.56396 & 0.196941 & 0.682223 &-0.417774\\
            $4$&0.56396 &-0.689293 & 0.170556 &-0.417774\\
            $5$&0.61778 & 0.499393 & 0        &-0.353124\\
            $6$&0.63680 &-0.013411 &-0.116145 &-0.47416\\
            $7$&0.66728 & 0        & 0        &-0.420338\\
            $8$&0.68492 &-0.372671 & 0.161371 &-0.22399\\
            $9$&1.03132 &-0.251613 &-0.251613 &-0.215682\\
            $10$&1.03132 &0.505446 &-0.135434 &0.257915\\
            $11$&1.54291 &-0.137190 &0.036760 &0.122291\\
            $12$&2.16954&-0.017784& -0.007206& -0.034589\\
			$13$&2.33189& 0.013514 &-0.003004 &-0.028353\\
            $14$&2.35773 &-0.041094& -0.011011 &0.027203\\
			$15$&2.45435 &-0.001277 &0.004767 &-0.025882\\
        \end{tabular}
        \caption{Coupling constants between the NV-center qubit and nuclear spins for the randomly generated realization of the environment used in all figures. The second column contains distances between the qubit and each nucleus.}
        \label{tab:couplingConstants}
    \end{ruledtabular}
\end{table}

\section{The system and the Hamiltonian} \label{sec:sys&H}

In this section, we provide a brief recapitulation of the Hamiltonian describing the NV-center spin qubit and the nuclear environment, as well as the resulting qubit-environment evolution in a formulation which is particularly well suited for the description of pure decoherence. 

The lowest energy states of the NV center constitute an effective spin-triplet \cite{XingFei1993}. In the presence of the magnetic field, the level spacing of the $S = 1$ states is uneven due to the zero-field splitting, which allows any two of the three states to be efficiently accessed. The usual choice of qubit employs the $m = 0, 1$ states, and we will restrict our studies to this subspace. The qubit Hamiltonian is thus given by $\hat{H}_{\mathrm{Q}} =(\Delta -\gamma_{\mathrm{e}}B_{\mathrm{z}}) \ket{1}\bra{1}$. Here, $\Delta = 2.87$~ GHz is the value of zero-field splitting and $\gamma_{\mathrm{e}} = 28.08$~MHz/T is the electron gyromagnetic ratio. We assumed the application of the magnetic field in the $z$ direction. As an environment, we are considering nuclear spins of the ${}^{13}C$ carbon isotope, which are known to randomly occur in the diamond crystal and are the main source of spin qubit decoherence in the best available samples \cite{Zhao2012, Kwiatkowski2018, Markham2011}. They are described by the Hamiltonian
$\hat{H}_{\mathrm{E}} = \gamma_{\mathrm{n}}B_{\mathrm{z}}\sum_{k=0}^{N-1}\hat{I}^z_k$, where $\gamma_{\mathrm{n}} = 10.71$~MHz/T is the nuclear gyromagnetic ratio of the ${}^{13}C$ atoms and $\hat{I}_k$ denotes the spin operator for nucleus $k$.

The hyperfine interaction between the qubit and the environment \cite{Smeltzer2011} is given by $\hat{H}_{\mathrm{QE}}=\ketbra{1}{1}\otimes\hat{V}$, where the effect that the qubit in the state $\ket{1}$ has on the environment is described by the operator
\begin{equation}
\hat{V}=\sum_{k=0}^{N-1}\sum_{j\in(\mathrm{x,y,z})}\!\!\!\mathbb{A}_k^{\mathrm{z},j}\hat{I}_k^j,
\end{equation}
while the state $\ket{0}$ is completely decoupled. 
The coupling constants which are dependent on the particular locations of the spinful carbon isotopes in a given sample are given by \cite{Gali2008}
\begin{equation}
    \mathbb{A}_k^{\mathrm{z},j} = \frac{\mu_0}{4\pi}\frac{\gamma_{\mathrm{e}}\gamma_{\mathrm{n}}}{r_k^3}\left(1 - 3\frac{(\textbf{r}_k\cdot\hat{\textbf{j}})(\textbf{r}_k\cdot\hat{\textbf{z}})}{r_k^2}\right).
    \label{eq:coupling_constants}
\end{equation}
Here, $\textbf{r}_k$ is the displacement vector between the NV-center and the $k$-th nuclear spin, $\hat{\textbf{j}}$ are versors in the three distinct directions, and $\mu_0$ is the magnetic permeability of the vacuum. 

We study initial states of the environment with no correlations between individual nuclei, $\hat{\rho}_{\mathrm{E}}(0) = \bigotimes_k\hat{\rho}_k(0)$,
where the density matrix of each nuclear spin is given by
\begin{equation}
    \hat{\rho}_k = \frac{1}{2}(\mathbb{1} + 2 p_k \hat{I}_k^{\mathrm{z}}).
    \label{rhok}
\end{equation}
Here, $p_k \in [-1;1]$ is the polarization of the $k$-th nucleus, which does not have to be the same for all nuclear spins.
   
For an NV-center spin qubit initially prepared in a superposition state, $|\psi\rangle=\alpha\ket{0} +\beta\ket{1}$, the qubit-environment state at time $t$ can be written in the form \cite{Roszak2018}
    \begin{equation}
        \hat{\rho}_{\mathrm{QE}}(t) =
        \begin{pmatrix}
            |\alpha|^2\hat{\rho}^{\mathrm{E}}_{00}(t) & \alpha\beta^*\hat{\rho}^{\mathrm{E}}_{01}(t)\\[5pt]
            \alpha^*\beta\hat{\rho}^{\mathrm{E}}_{10}(t) & |\beta|^2\hat{\rho}^{\mathrm{E}}_{11}(t)
        \end{pmatrix},
        \label{sigma}
    \end{equation}
    where the environmental matrices are given by
    \begin{equation}
        \label{rij}
        \hat{\rho}^{\mathrm{E}}_{ij}(t) = \hat{w}_i(t)\hat{\rho}_{\mathrm{E}}(0)\hat{w}_j^\dagger(t).
    \end{equation}
The environmental evolution operators, $\hat{w}_i(t)$, which describe the evolution of the  conditional on the pointer state of the qubit, either $\ket{0}$ or $\ket{1}$, are given by $\hat{w}_0(t)=\exp{[-i\hat{H}_{\mathrm{E}} t]}$ and $\hat{w}_1(t)=\exp{[-i(\hat{V}+\hat{H}_{\mathrm{E}}) t]}$.


\section{The scheme\label{sec:ent}}

\begin{figure}[tb!]
    \centering
    \includegraphics[width=1\columnwidth]{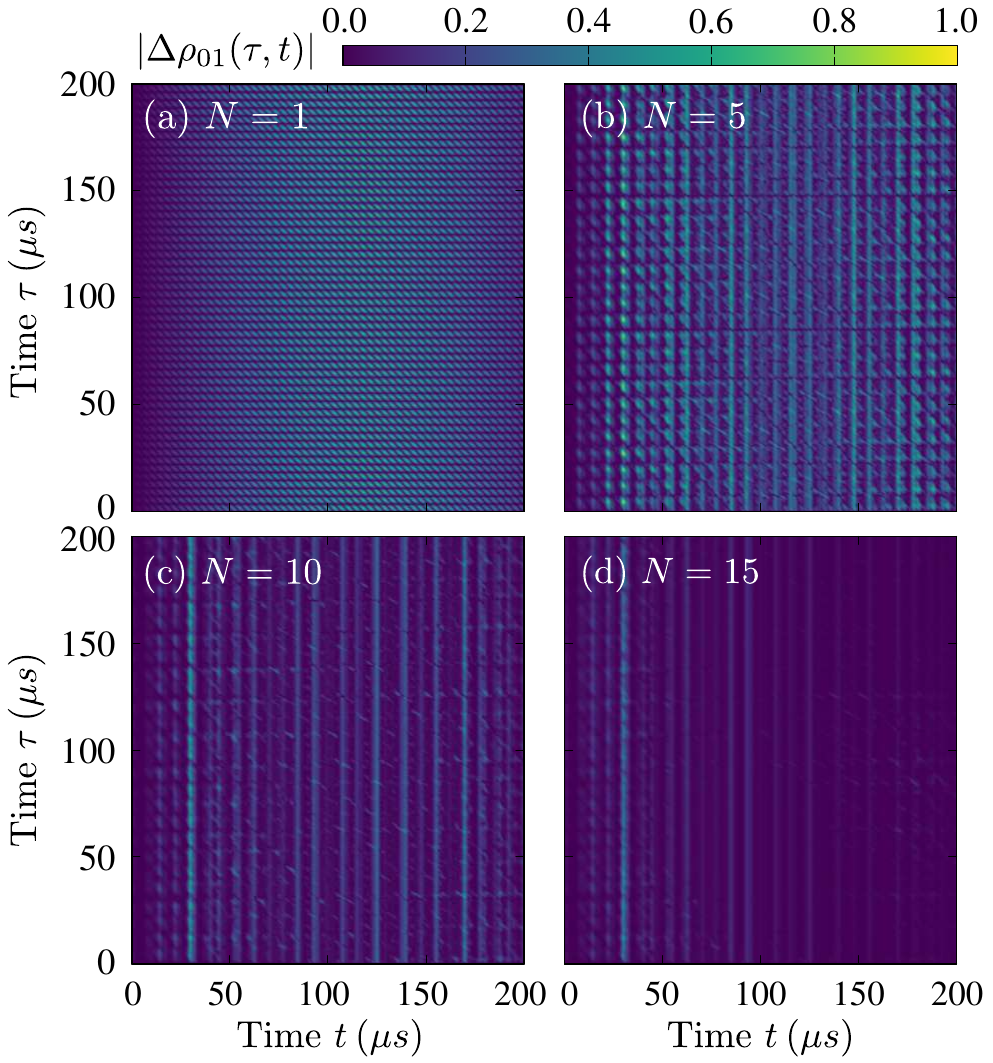}
    \caption{The absolute value of the signal $\Delta\rho_{01}(\tau, t)$ as a function of both times $t$ and $\tau$ for: (a) $N=1$ spin, (b) $N=5$ spins, (c) $N=10$ spins, and (d) $N=15$ spins. We choose $p_k = 1$ for all $k$ and $B_{\mathrm{z}} = 25$~G.}
    \label{fig:coherenceDiff}
\end{figure}

The scheme introduced in Ref.~\cite{Roszak2019} is meant to facilitate the transfer of information about the evolution of the environment conditional on the pointer state of the qubit into the qubit evolution. To this end, it uses a preparation phase in which the qubit is initialized in one of the pointer states, $|0\rangle$ or $|1\rangle$, and the qubit and environment evolve jointly for a controlled time $\tau$. Since their interaction can only lead to pure decoherence of the qubit, the state of the qubit is not affected during this stage, but the environment does evolve according to
\begin{subequations}
\begin{eqnarray}
    \ketbra{0}{0}\otimes \hat{\rho}_{\mathrm{E}}(0)&\rightarrow&\ketbra{0}{0}\otimes \hat{\rho}_{00}^{\mathrm{E}}(\tau),\\[3pt]
    \ketbra{1}{1}\otimes \hat{\rho}_{\mathrm{E}}(0)&\rightarrow&\ketbra{1}{1}\otimes \hat{\rho}_{11}^{\mathrm{E}}(\tau),
\end{eqnarray}    
\end{subequations}
as evident from \eqnref{sigma} with $\beta=0$ and $\alpha=0$, respectively. 

At time $\tau$, the qubit is excited into the equal superposition state, $|+\rangle =(|0\rangle+|1\rangle)/\sqrt{2}$, and now the qubit-environment interaction leads to decay of the qubit coherence. Since the environment is in a new initial state after the preparation stage, which is eiter $\hat{\rho}_{00}^{\mathrm{E}}(\tau)$ or $\hat{\rho}_{11}^{\mathrm{E}}(\tau)$, the qubit coherence at time $t$, elapsed after preparation of state $|+\rangle$, evolves according to 
\begin{equation}
    \label{rho01}
    \rho_{01}^{(i)}(\tau, t) = \frac{1}{2}\Tr\left(\hat{w}_0(t)\hat{\rho}_{ii}^{\mathrm{E}}(\tau)\hat{w}_1^\dagger(t)\right),
\end{equation}
with $i=0,1$.

If the interaction is not entangling for a given initial state of the environment, $\hat{\rho}_E(0)$, then $\hat{\rho}_{00}^{\mathrm{E}}(\tau)=\hat{\rho}_{11}^{\mathrm{E}}(\tau)$ at all $\tau$ \cite{Roszak2015} and the evolution of the qubit coherence given by \eqnref{rho01} does not depend on the preparation phase. No differences in the evolution will be observed also for commuting operators $\hat{w}_0(t)$ and $\hat{w}_1(t')$. Otherwise, differences between $\rho_{01}^{(0)}(\tau, t)$ and $\rho_{01}^{(1)}(\tau, t)$ are observed at almost all times $t$ \cite{Roszak2021}.

For the NV-center spin qubit and the initial environmental states under study, \eqnref{rhok}, we have $[\hat{\rho}_{\mathrm{E}}(0),\hat{w}_0(\tau)]=0$. This means that  the preparation phase with qubit in state $\ket{0}$ has no effect on the state of the environment, $\hat{\rho}_{00}^{\mathrm{E}}(\tau)=\hat{\rho}_{\mathrm{E}}(0)$. Thus, no preparation phase is necessary for the $\ket{0}$ pointer state, reducing this part of the scheme to a straightforward qubit-coherence measurement, $\rho_{01}^{(0)}(\tau, t) = \frac{1}{2}\Tr\left(\hat{w}_0(t)\hat{\rho}_{\mathrm{E}}(0)\hat{w}_1^\dagger(t)\right)$. 

Furthermore, the free Hamiltonian of the environment and the interaction do not commute, which translates into the non-commutation of $\hat{w}_0(t)$ and $\hat{w}_1^\dagger(t')$ at almost all instances of time. This means that the scheme can detect QEE for the system under study, and it can therefore be used to quantify the initial polarization of the environment.

In the studied system, the maximum amount of QEE that can be generated during the evolution is proportional to the initial polarization of the nuclear spins \cite{Strzalka2020}. Because the form of the interaction Hamiltonian is set and the initial states of the environment are limited to a given class of states given by \eqnref{rhok}, the difference in the coherence evolution, each given by \eqnref{rho01}, is proportional to the amount of generated entanglement and thus contains information about the initial polarization of the environment.

We work in the rotating frame, in order to neglect the qubit phase accumulation due to the well-known and well-controlled energy splitting. The coherence, $\rho_{01}^{(0/1)}(\tau, t)$, can be expressed as $\prod_k L_k^{(0/1)}(\tau, t)$, where $L_k^{(0/1)}(\tau, t)$ is the coherence that would be measured if the environment consisted of only the $k$-th nuclear spin. The difference of the two signals $\Delta\rho_{01}(\tau, t) = \rho_{01}^{(0)}(\tau, t) - \rho_{01}^{(1)}(\tau, t)$, which contains the information about initial environmental polarization, is expressed as 
\begin{equation}
    \Delta\rho_{01}(\tau, t) = \frac{1}{2}\prod_{k=0}^{N-1}L_k^{(0)}(t) - \frac{1}{2}\prod_{k=0}^{N-1}L_k^{(1)}(\tau, t).
    \label{eq:signalDifference}
\end{equation}
Here $L_k^{(0)}(t)$ and $L_k^{(1)}(\tau, t)$ are given by
\begin{subequations}
    \begin{eqnarray}
	L_k^{(0)}(t)&=& \left(\frac{\omega+\mathbb{A}_k^{\mathrm{z},\mathrm{z}}}{\omega_{k}}\sin\frac{\omega t}{2}\sin\frac{\omega_{k}t}{2} + \cos\frac{\omega t}{2}\cos\frac{\omega_{k}t}{2}\right)\\
             \nonumber
        &&+ ip_k\left(\frac{\omega+\mathbb{A}_k^{\mathrm{z},\mathrm{z}}}{\omega_{k}}\cos\frac{\omega t}{2}\sin\frac{\omega_{k}t}{2}-\sin\frac{\omega t}{2}\cos\frac{\omega_{k}t}{2}\right), \\
	            L_k^{(1)}(\tau, t) &=& L_k^{(0)}(t) -ip_kC_k(\tau, t),
    \end{eqnarray}
\end{subequations}
with
\begin{equation}
    C_k(\tau, t)=-2\frac{(\mathbb{A}_k^{\mathrm{z}, \mathrm{x}})^2 + (\mathbb{A}_k^{\mathrm{z}, \mathrm{y}})^2}{\omega_{k}^2}\sin\frac{\omega_{k}\tau}{2}\sin\frac{\omega t}{2}\sin\frac{\omega_{k}(\tau+t)}{2}.
\end{equation}
The frequency $\omega = \gamma_{\mathrm{n}}B_{\mathrm{z}}$ is independent of the particular nuclear spin $k$, and is well known and controllable, while the frequencies $\omega_{k} = \sqrt{(\mathbb{A}_k^{\mathrm{z}, \mathrm{x}})^2 + (\mathbb{A}_k^{\mathrm{z}, \mathrm{y}})^2 + (\omega + \mathbb{A}_k^{\mathrm{z},\mathrm{z}})^2}$ strongly depend on the location of the spin $k$ (see Table \ref{tab:couplingConstants}). 

All terms in \eqnref{eq:signalDifference} that are not multiplied by $C_k(\tau, t)$ cancel out (the derivation is given in Appendix \ref{sec:induction}), which allows it to be rewritten in the form
    \begin{equation}
        \Delta\rho_{01}(\tau, t) = \frac{1}{2}\sum_{m=0}^{N-1} ip_m C_m(\tau, t)\prod_{n=0}^{m-1} L_n^{(0)}(t)\prod_{k=m+1}^{N-1} L_k^{(1)}(\tau, t).
    \label{eq:deltaRhoGeneral}
    \end{equation}
This form is more concise and allows for the following analysis and extraction of the lower bounds of the average nuclear polarization from the maximum measured difference in the evolution of qubit coherence. At this stage no approximations have been made. 
	
    
\section{Estimation of polarization}
\label{sec:results}


\begin{figure}[tb!]
    \centering
    \includegraphics[width=1\columnwidth]{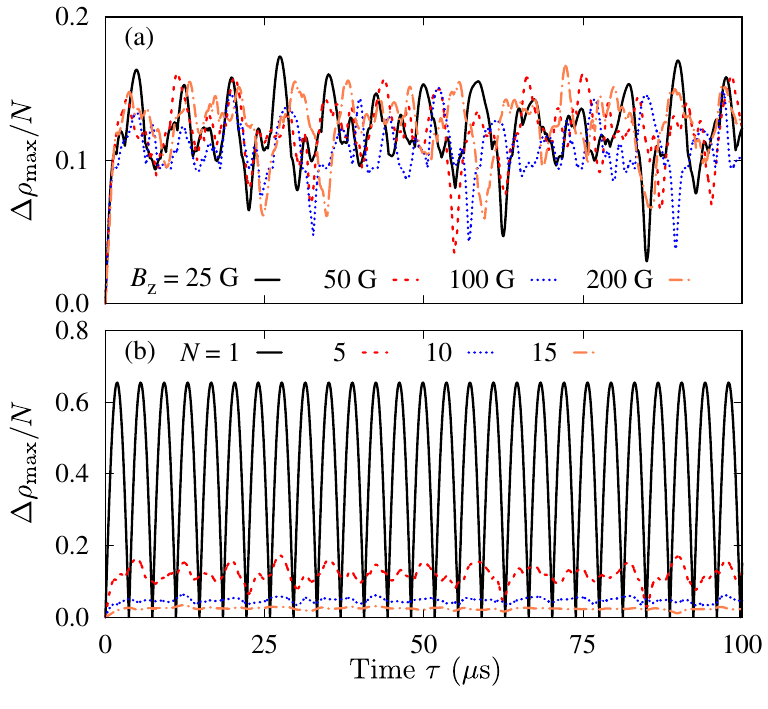}
    \caption{Estimation of the minimal value of the polarization for the time-independent case as a function of time $\tau$. Panel (a): polarization value for $N=5$ nuclear spins for different values of the magnetic field. Panel (b): polarization value in the magnetic field $B_{\mathrm{z}} = 25$~G for different numbers of nuclear spins. We choose $p_k = 1$ for all $k$.}
    \label{fig:timeIndependentEstimation}
\end{figure}

\subsection{Time time-independent approach}
To find a useful estimation of the average nuclear polarization, we work with the absolute value of the difference of coherence. We start from \eqnref{eq:deltaRhoGeneral} and using the triangle inequality we get
    \begin{equation}
        \lvert\Delta\rho_{01}(\tau, t)\rvert \!\leq \!\frac{1}{2}\!\sum_{m=0}^{N-1} \left\lvert p_m\right\rvert\left\lvert C_m(\tau, t)\right\rvert\!\prod_{n=0}^{m-1} \!\left\lvert L_n^{(0)}(t)\right\rvert\!\prod_{k=m+1}^{N-1} \!\left\lvert L_k^{(1)}(t)\right\rvert.
        \label{eq:delta_rho_triangle_ineq}
    \end{equation}
At this point, we only need to bind $\lvert C_k(\tau, t)\rvert$ and $\lvert L_k^{(0/1)}(t)\rvert$ to extract the value of the polarization and remove the dependence on the times and frequencies. Since $\omega_k\ge\omega+\mathbb{A}_k^{\mathrm{z},\mathrm{z}}$ and $|p_k|\le 1$, we get
\begin{eqnarray}
    \left\lvert L_k^{(i)}(t)\right\rvert &\leq& 1,
    \label{eq:ineq_Lk}
\end{eqnarray}
for all $k$ and $i=0,1$.
Since the function $L_k^{(i)}(t)$ governs the decoherence that would be induced by a single environmental spin, the above limitation follows from the definition of the qubit density matrix. Analogously, since the amplitude of $\lvert C_k(\tau, t)\rvert$ is bound by 
\begin{equation}
\label{dzielone}
    \frac{(\mathbb{A}_k^{\mathrm{z}, \mathrm{x}})^2 + (\mathbb{A}_k^{\mathrm{z}, \mathrm{y}})^2}{\omega_{k}^2} \leq 1,
\end{equation}
we get
\begin{equation}
    \left\lvert C_k(\tau, t)\right\rvert \leq 2,
    \label{eq:ineq_Ck}
\end{equation}
which is satisfied for every $k$. 

This leads to an extremely simple form of the lower bound for the difference of coherence,
\begin{equation}
    \left\lvert\Delta\rho_{01}(\tau, t)\right\rvert \leq \sum_{k=0}^{N-1}|p_k|,
\end{equation}
which is directly related to the sum of the amplitudes of the  polarizations of spinful nuclei. This can be used to estimate the average value. We introduce the quantity $\Bar{p}$ for the average polarization $\Bar{p} = \sum_{k=0}^{N-1} |p_k| / N$. Although the signal still depends on the times $t$ and $\tau$, we do not need to know their values. The significant quality here is the maximum of the function $|\Delta\rho_{01}(\tau, t)|$, which we denote as $\Delta\rho_{\mathrm{max}}$. This allows us to bind the coherence signal with it, leading to the relation which estimates the average polarization,
\begin{equation}
    \Bar{p} \geq \frac{\Delta\rho_{\mathrm{max}}}{N}.
\end{equation}

To find how accurate this crude estimation is, we performed numerical simulations for different numbers of nuclear spins and different values of the magnetic field for a fully polarized environment ($p_k=1$ for all $k$). The coupling constants from Table \ref{tab:couplingConstants} are used regardless of the number of spins taken into account, in such a way that $k=1,\dots,N$.  
In \subfigref{fig:timeIndependentEstimation}{a} we present the estimated value of the polarization as a function of $\tau$ for $N=5$ nuclear spins. The results show that the average polarization of the environment is at least on the order of $0.17$, and do not predict strong magnetic field dependence on this protocol. In \subfigref{fig:timeIndependentEstimation}{b} we show how different numbers of ${}^{13}C$ nuclei affect the final result. Even for an NV center that interacts with a single nuclear spin ($N=1$) we do not reach unity,
because the amplitude of the observed oscillation is proportional to the ratio of perpendicular coupling constants with respect 
to the frequency $\omega_k$ for $k=1$, \eqnref{dzielone}. However, the signal never vanishes and even for $N=15$ (where almost half of the spins interact weakly), we are able to state if the environment was initially polarized.
	

\begin{figure}[tb!]
    \centering
    \includegraphics[width=1\columnwidth]{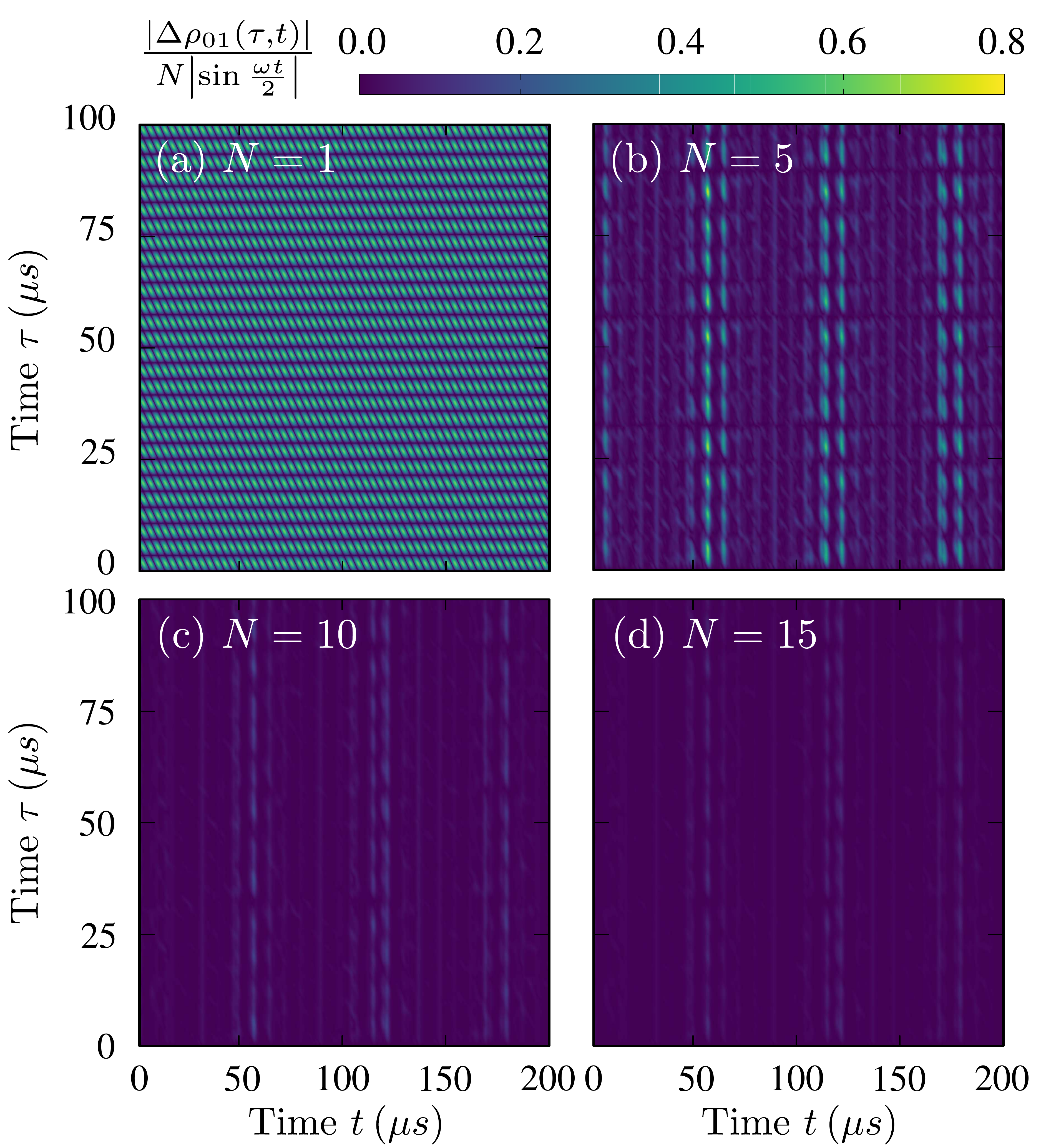}
    \caption{Estimation of the minimal value of the polarization for the time-dependent scheme as a function of both times $\tau$ and $t$ for: (a) $N = 1$ spin, (b) $N=5$ spins, (c) $N=10$ spins, and (d) $N=15$ spins. We choose $p_k = 1$ for all $k$ and $B_{\mathrm{z}} = 100$~G.}
    \label{fig:timeDependentN}
\end{figure}

\begin{figure}[tb!]
    \centering
    \includegraphics[width=1\columnwidth]{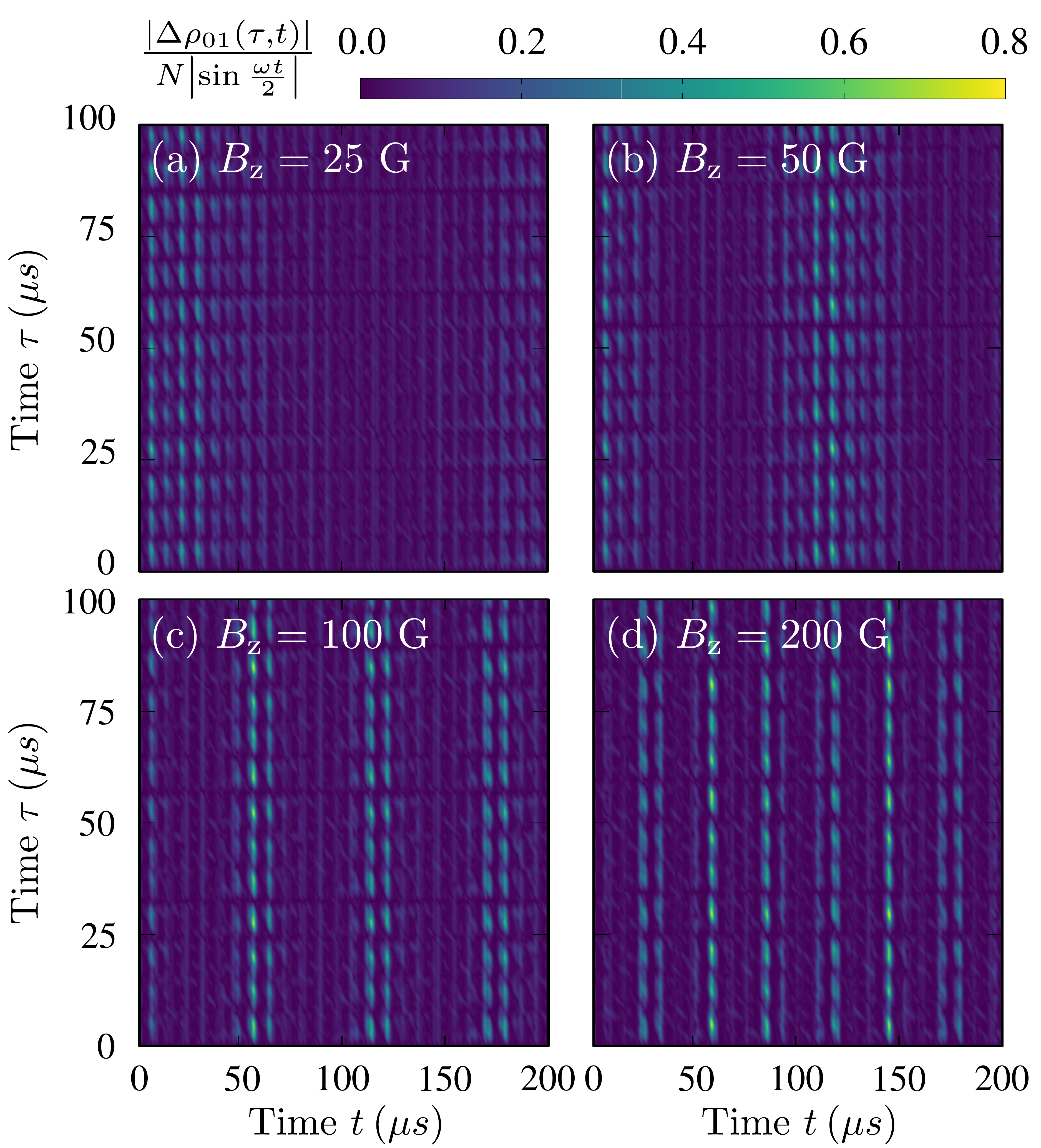}
    \caption{Estimation of the minimal value of the polarization for the time-dependent scheme as a function of both times $\tau$ and $t$ for: (a) $B_{\mathrm{z}} = 25$~G, (b) $B_{\mathrm{z}} = 50$~G, (c) $B_{\mathrm{z}} = 100$~G, and (d) $B_{\mathrm{z}} = 200$~G. We choose $p_k = 1$ for all $k$ and $N = 5$ spins.}
    \label{fig:timeDependentB}
\end{figure}

\subsection{Time $t$ dependent estimation of polarization}

Although the dependence of the results on the ratio of perpendicular coupling constants with respect to the frequency $\omega_k$ for each nuclear spin, which is the main factor that hinders the faithfulness
of obtained lower bound with respect to the actual mean value of the initial environmental polarization cannot be helped, there is a straightforward way to increase the precision of the proposed method. 
Since the applied magnetic field is known, and it is relatively easy to keep track of the time $t$ that elapsed after the excitation of the superposition of the qubit, the limit on the function $C_k(\tau, t)$
can be improved using the factor $\sin(\omega t/2)$,
\begin{equation}
    \left\lvert C_k(\tau, t)\right\rvert \leq 2\left\lvert\sin\frac{\omega t}{2}\right\rvert.
\end{equation}
Thus the lower bound on the average polarization is given by,
\begin{equation}
    \Bar{p} \geq \frac{\left\lvert\Delta\rho_{01}(\tau, t)\right\rvert}{N\left\lvert\sin\frac{\omega t}{2}\right\rvert}.
\end{equation}

\ifigref{fig:timeDependentN} presents the results for fully polarized spins ($p_k = 1$ for all $k$) in a magnetic field $B_{\mathrm{z}} = 100$~G. For an environment of a single spin, $N=1$, [\subfigref{fig:timeDependentN}{a}] we extract the value of polarization $\Bar{p} \geq 0.62$, which is a similar result to the time $t$ independent case result. However, the advantage of time-dependent variation of this scheme is already visible for an environment of five spins, $N=5$, [\subfigref{fig:timeDependentN}{b}]. Here, we obtain $\Bar{p} \geq 0.73$, while the time-independent estimation gives us $\Bar{p} \geq 0.17$.

Although the accuracy of this estimation decays with larger environments, the difference in the lower bound of polarization between $10$ nuclei, \subfigref{fig:timeDependentN}{c}, and $15$ nuclei, \subfigref{fig:timeDependentN}{d}, is relatively small and the lower bound on the polarization is $0.19$ and $0.10$, respectively. This is due to the fact that the last $5$ spins are weakly coupled to the NV center, so they do not significantly impact the qubit coherence evolution. Note that although the measurable lower bound on the average nuclear polarization is much smaller than its actual value, there is still a large improvement obtained over the time-independent method (which
yielded $0.0632$ for $N=10$ and $0.0338$ for $N=15$).

In \figref{fig:timeDependentB} we show the influence of the magnetic field on the simulated measurement outcomes for $N=5$ nuclear spins. Similarly as in the time-independent variant of the protocol, the magnetic field-dependence is not significant. The lower bound of the polarization varies moderately between different panels, yielding $\Bar{p} \geq 0.55$, $\Bar{p} \geq 0.64$, $\Bar{p} \geq 0.73$, $\Bar{p} \geq 0.74$, for $B_{\mathrm{z}} = 25$~G, $B_{\mathrm{z}} = 50$~G, $B_{\mathrm{z}} = 100$~G, and  $B_{\mathrm{z}} = 200$~G, respectively.

\begin{figure}[tb!]
    \centering
    \includegraphics[width=1\columnwidth]{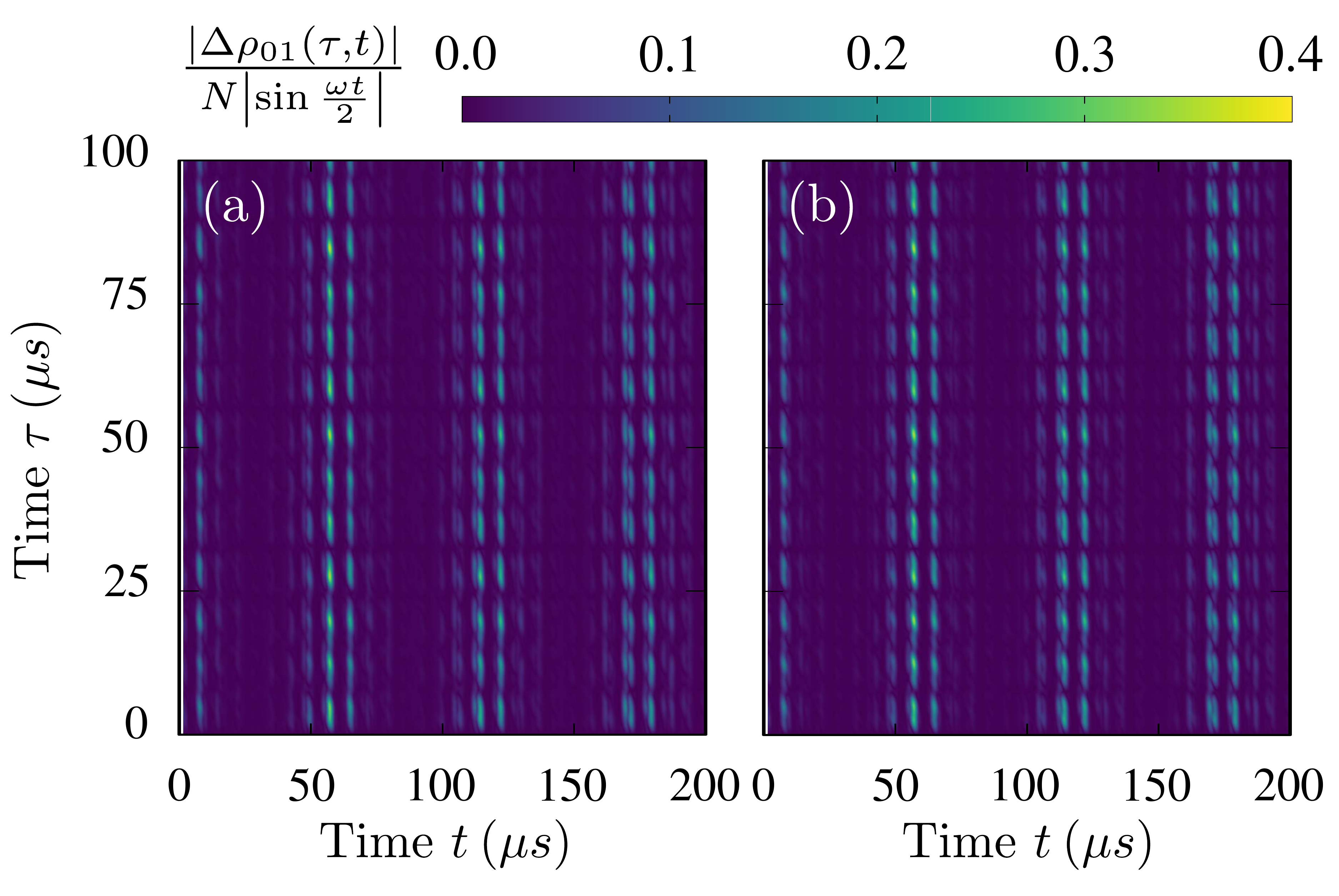}
    \caption{Estimation of the minimal value of the polarization for the time-dependent scheme as a function of both times $\tau$ and $t$ for $\Bar{p} = 0.5$, $B_{\mathrm{z}} = 100$~G, and $N=5$ where: (a) $p_k = 0.5$ for all $k$; (b) $p_k$ is different for all $k$, and spins that are closer to the NV center are polarized stronger.}
    \label{fig:timeDependentRandom}
\end{figure}

\begin{figure}[tb!]
    \centering
    \includegraphics[width=1\columnwidth]{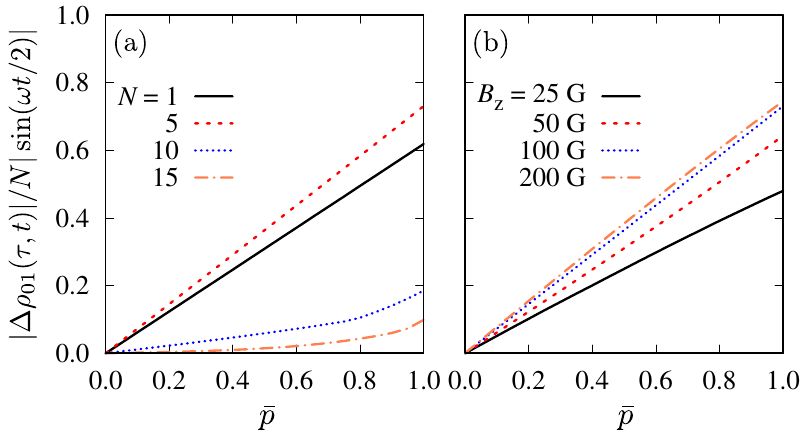}
    \caption{Estimation of the minimal value of the polarization for the time-dependent scheme as a function of real average polarization $\Bar{p}$ for (a) different number of nuclei in a magnetic field $B_{\mathrm{z}} = 100$~G; (b) different magnetic field for $N=5$.}
    \label{fig:pVsp}
\end{figure}

In the following, we study an environment that is only partially polarized due to an unideal polarization protocol. In \figref{fig:timeDependentRandom} we present plots for five nuclei in a magnetic field $B_{\mathrm{z}} = 100$~G that have an average value of polarization $\Bar{p} = 0.5$. In \subfigref{fig:timeDependentRandom}{a}, a uniform nuclear polarization is assumed. In contrast, \subfigref{fig:timeDependentRandom}{b} considers a spatially non-uniform polarization distribution characterized by a mean $\Bar{p}=0.5$ and standard deviation $\sigma_p=0.261$, with polarization strength increasing for nuclei located closer to the defect center. The observed signal is similar in both cases and the obtained estimate for the average spin is $\Bar{p}\geq 0.36$ when all polarizations are the same, and $\Bar{p}\geq 0.35$, for different polarizations. The two values are comparable, which is in line with our expectations, since the value of average polarization does not change the position of maxima.

Lastly, the estimate as a function of the actual average polarization of the environment (with constant $p_k$) are plotted in \subfigref{fig:pVsp}{a} for different numbers of ${}^{13}C$ isotopes at magnetic field
$B_{\mathrm{z}} = 100$~G, and in \subfigref{fig:pVsp}{b} for different magnetic field for $N=5$ nuclei. The dependence is linear for small numbers of nuclei and moderately high magnetic fields, so that the measured lower bound is directly proportional to the average value of the polarization for the whole range of polarizations. Once the number of environmental nuclei is larger, the estimation performs better for large polarizations. Incidentally, if the applied magnetic field is too low, then there is a slight under-performance of the estimation for higher polarizations,
which suggests that the protocol should not be used at very low magnetic fields.

\section{Conclusion \label{sec:conclusion}}

We have introduced a method to find the lower bound on the initial polarizations of spinful nuclei in ${}^{13}C$ carbon isotopes within the diamond lattice. This involves analyzing the difference of the coherence evolution of an NV center spin qubit following a controlled period of time when the qubit was prepared in different spin states. Because of the numerous unknown variables characterizing the interaction with the nuclear environment in practical scenarios and the intricate nature of the decoherence function resulting from the presence of even a few carbon isotope atoms, directly determining the polarization is not straightforward. Thus, we derive two inequalities for time-independent and time-dependent scenarios, which allow estimating the initial value of this polarization by simply measuring the difference of the coherence evolution for qubit spin states. 

We demonstrated the functionality of the method using a setting in which multiple nuclear spins were randomly positioned near the qubit within the diamond crystal lattice. Given the short-range interaction between the qubit and its surrounding environment, the experimental setup is not expected to involve a significantly larger number of nuclear spins making noticeable contributions.

We find that the time-dependent method (which requires taking into account the time at which a given level of decoherence is measured) is significantly better than the time-independent one, and it gives quite accurate prediction even for a small magnetic field. The advantage of the suggested approach lies in the simplicity of the necessary measurements, as they do not require direct access to any degrees of freedom of the environment. Instead, all information regarding the environment's state is obtained through the transfer of information into the qubit state during the collective evolution of the system and the environment.

\section{Acknowledgement}
The contribution of K.R.~was supported within the QuantERA II Programme that has received funding from the EU H2020 research and innovation programme
under Grant Agreement No 101017733, and with funding organisation MEYS (The Ministry of Education, Youth and Sports) of the Czech Republic.

\appendix
\onecolumngrid
\section{Qubit coherence evolution \label{sec:induction}}

The coherence evolution of the qubit initially prepared in state $\ket{1}$, interacting with $N$ nuclear spins, evolves according to
\begin{equation}
    \rho_{01}^{(1)}(\tau, t) = \frac{1}{2}\prod_{k=0}^{N-1}L_{k}^{(1)}(\tau, t) = \frac{1}{2}\prod_{k=0}^{N-1}\left[L_{k}^{(0)}(t) - ip_{k}C_{k}(\tau, t)\right].
    \label{eq:rho1signal}
\end{equation}
\textit{Proposition:}$\forall N \in \mathbb{N}^+$ hold
\begin{equation}
    \rho_{01}^{(1)}(\tau, t) =\frac{1}{2}\prod_{k=0}^{N-1}L_{k}^{(0)}(t) - \frac{1}{2}i\sum_{m=0}^{N-1}p_{m}C_{m}(\tau, t)\prod_{n=0}^{m-1}L_{n}^{(0)}(t)\prod_{k=m+1}^{N-1}L_{k}^{(1)}(\tau, t).
    \label{eq:proposition}
\end{equation}
\textit{Proof:} For base case $N=1$ we get
\begin{equation}
    \rho_{01}^{(1)}(\tau, t) =\frac{1}{2}L_{0}^{(0)}(t) - \frac{1}{2}ip_{0}C_{0}(\tau, t) = \frac{1}{2}L_{0}^{(1)}(\tau, t),
\end{equation}
which agrees with the result from \eqnref{eq:rho1signal}. Now, by taking $K = N+1$, we obtain
\begin{equation}
    \begin{aligned}
        2\rho_{01}^{(1)}(\tau, t)\! =\prod_{k=0}^{N}\left[L_{k}^{(0)}(t) - ip_{k}C_{k}(\tau, t)\right] = \left(L_{N}^{(0)}(t) - ip_{N}C_{N}(\tau, t)\right)\!\prod_{k=0}^{N-1}\left[L_{k}^{(0)}(t) - ip_{k}C_{k}(\tau, t)\right].
    \end{aligned}
\end{equation}
Let us assume that \eqnref{eq:proposition} holds. Then
\begin{equation}
    \begin{aligned}
        2\rho_{01}^{(1)}(\tau, t) = L_{N}^{(0)}(t)\prod_{k=0}^{N-1}L_{k}^{(0)}(t) - ip_{N}C_{N}(\tau, t)\prod_{k=0}^{N-1}L_{k}^{(0)}(t) - iL_{N}^{(1)}(\tau, t)\sum_{m=0}^{N-1}p_{m}C_{m}(\tau, t)\prod_{n=0}^{m-1}L_{n}^{(0)}(t)\prod_{k=m+1}^{N-1}L_{k}^{(1)}(\tau, t).
    \end{aligned}
\end{equation}
In the first and last terms, we can incorporate signals $L_{N}^{(0)}(t)$ and $L_{N}^{(1)}(\tau, t)$ into the product, respectively
\begin{equation}
    \begin{aligned}
        2\rho_{01}^{(1)}(\tau, t) = \prod_{k=0}^{N}L_{k}^{(0)}(t) - ip_{N}C_{N}(\tau, t)\prod_{k=0}^{N-1}L_{k}^{(0)}(t) - i\sum_{m=0}^{N-1}p_{m}C_{m}(\tau, t)\prod_{n=0}^{m-1}L_{n}^{(0)}(t)\prod_{k=m+1}^{N}L_{k}^{(1)}(\tau, t).
    \end{aligned}
\end{equation}
Now, the second term already fits the last summation, leaving us with
\begin{equation}
    \begin{aligned}
        2\rho_{01}^{(1)}(\tau, t) = \prod_{k=0}^{N}L_{k}^{(0)}(t) - i\sum_{m=0}^{N}p_{m}C_{m}(\tau, t)\prod_{n=0}^{m-1}L_{n}^{(0)}(t)\prod_{k=m+1}^{N}L_{k}^{(1)}(\tau, t),
    \end{aligned}
\end{equation}
which ends the induction.


\twocolumngrid
\bibliography{bibliography,quantum19, projectBib}

\end{document}